\documentstyle[12pt]{article}
\renewcommand{\thefootnote}{\fnsymbol{footnote}}

\catcode`\@=11
\def\maketitle{\par
 \begingroup
 \def\thefootnote{\fnsymbol{footnote}}
 \def\@makefnmark{\hbox
 to 0pt{$^{\@thefnmark}$\hss}}
 \if@twocolumn
 \twocolumn[\@maketitle]
 \else \newpage
 \global\@topnum\z@ \@maketitle \fi\thispagestyle{empty}\@thanks
 \endgroup
 \setcounter{footnote}{0}
 \let\maketitle\relax
 \let\@maketitle\relax
 \gdef\@thanks{}\gdef\@author{}\gdef\@title{}\let\thanks\relax}
\def\@maketitle{\newpage
 \null
 \hbox to\textwidth{\hfil\hbox{\begin{tabular}{r}\@preprint\end{tabular}}}
 \vskip 2em \begin{center}
 {\Large\bf \@title \par} \vskip 1.5em {\normalsize \lineskip .5em
\begin{tabular}[t]{c}\@author
 \end{tabular}\par}
 \end{center}
 \par
 \vskip 1.5em}
\def\preprint#1{\gdef\@preprint{#1}}
\def\abstract{\if@twocolumn
\section*{Abstract}
\else \normalsize
\begin{center}
{\large\bf Abstract\vspace{-.5em}\vspace{0pt}}
\end{center}
\quotation
\fi}
\def\endabstract{\if@twocolumn\else\endquotation\fi}
\catcode`\@=12
\begin{document}
\baselineskip=.285in

\preprint{SNUTP/97-022 \\[-2mm] gr-qc/9707011}

\title{\Large\bf Charged Black Cosmic String
\protect\\[1mm]\ }
\author{{\normalsize Nakwoo Kim, Yoonbai Kim${}^{\ast}$ and Kyoungtae Kimm}\\
{\normalsize\it Department of Physics and Center for Theoretical Physics,}\\
{\normalsize\it Seoul National University, Seoul 151-742, Korea}\\
{\normalsize\it nakwoo$@$phya.snu.ac.kr, dragon$@$phya.snu.ac.kr}\\
{\normalsize\it ${}^{\ast}$Department of Physics, Sung Kyun Kwan University, 
Suwon 440-746, Korea}\\
{\normalsize\it yoonbai$@$cosmos.skku.ac.kr}}
\date{}
\maketitle

\begin{center}
{\large\bf Abstract}\\[3mm]
\end{center}
\indent\indent
Global $U(1)$ strings 
with cylindrical symmetry are studied in anti-de Sitter
spacetime. According as the magnitude of negative cosmological constant, they
form regular global cosmic strings, extremal black cosmic strings and charged 
black cosmic strings, but no curvature singularity is involved. 
The relationship between the topological charge of a neutral global string 
and the black hole charge is clarified by duality transformation.
Physical relevance as straight string is briefly discussed.
\vspace{1.5mm}

\noindent PACS number(s): 11.27.+d, 04.40-b, 04.70.Bw


\newpage

\pagenumbering{arabic}
\thispagestyle{plain}

Cosmic strings are viable extended objects 
in cosmology~\cite{VS}. 
A way to understand basic physical ingredients 
of cosmic strings is to study a straight string along an axis,
which reduces one spatial dimension.
Then, the (2+1) dimensional correspondence is the 
particle-like solitonic excitations so-called vortices in curved spacetime, 
and the conic space due to massive point source is enough for 
the description of asymptotic region outside the local vortex core.
Recently, black hole solutions have been reported in (2+1) dimensional
anti-de Sitter spacetime~\cite{BTZ} 
in addition to known hyperbolic solutions~\cite{DJ}, 
and these Ba\~{n}ados-Teitelboim-Zanelli (BTZ)
black hole solutions have been extensively studied in a variety
of models~\cite{Cle}.
Here we may raise a question that what are the string-like counterpart
of these BTZ black holes in cosmology.
Specifically, whether the vortices in anti-de Sitter space can 
constitute black holes in (2+1)D, or straight black strings 
in (3+1)D. 
The objects of our interest are global $U(1)$
vortices~\cite{HS,Gre}.
 
It has been shown in Ref.~\cite{Gre} that global $U(1)$
strings coupled to Einstein gravity with zero cosmological constant lead to
a physical curvature singularity. 
Then, how does the constant negative vacuum energy affect the 
global strings?
In this paper, we consider the effect of the negative cosmological
constant to the global $U(1)$ vortices in (2+1)D and find three types
of regular solutions of which base manifolds form 
(i) smooth hyperbola, (ii) extremal charged black hole and 
(iii) charged black hole with two horizons.
{}For all these static solutions, the physical singularity can be avoided,
which is different from the zero cosmological constant case. 
Suppose the magnitude of the negative cosmological constant is 
extremely small like the lower bound of it in the present universe,
$|\Lambda| \le  10^{-83}{\rm GeV}^2$.
Under this perfect toy environment with no fluctuation
the global string  may be born as a black string 
with large horizon size $r_H$ in the early universe, i.e.,
$r_H \sim 10^6{\rm pc}$ for the grand unification scale 
and $r_H\sim 10^{-2}{\rm A.U.}$ for the electroweak scale.

A cylindrically symmetric metric with boost invariance in the $z$-direction
can be written as
\begin{eqnarray}
\label{cyl}
ds^2=e^{2N(r)}B(r)(dt^2-dz^2)-\frac{dr^2}{B(r)}-r^2d\theta^2.
\end{eqnarray}
The physics is reduced to (2+1) dimensional one under this metric.
Another well-known (2+1)D static metric is written under conformal gauge:
\begin{eqnarray}
\label{conf}
ds^2=\Phi(R)dt^2-b(R)(dR^2+ R^2d\Theta^2). 
\end{eqnarray}
{}For a spinless point particle source of mass $m$ at the origin,
the general anti-de Sitter solution is
\begin{eqnarray}
b&=&\frac{4\varepsilon c^2}{
|\Lambda| R^2\Big[ (R/R_0)^{\sqrt{\varepsilon}c}
-(R_0/R)^{\sqrt{\varepsilon}c}
\Big]^2} \label{beq}
\\
\label{phieq}
\Phi&=&\sqrt{\varepsilon}
\frac{(R/R_0)^{\sqrt{\varepsilon}c}
+(R_0/R)^{\sqrt{\varepsilon}c}}{
(R/R_0)^{\sqrt{\varepsilon}c}
-(R_0/R)^{\sqrt{\varepsilon}c}},
\end{eqnarray}
where $\varepsilon$ is $\pm 1$ for $\Lambda <0$.
When $\varepsilon=+1$, a coordinate transformation 
\begin{eqnarray}
r=\frac{2}{|\Lambda|^{1/2}}
\frac{1}{|R^{(1-4Gm)}-R^{-(1-4Gm)}|} 
~~\mbox{and}~~\theta=(1-4Gm)\Theta 
~~(c=1-4Gm)
\end{eqnarray}
leads to
\begin{eqnarray}
ds^{2}=(1+|\Lambda|r^{2})dt^{2}
-\frac{dr^2}{1+|\Lambda|r^{2}}-r^{2}d\theta^{2}. 
\end{eqnarray}
It describes a hyperbola with deficit angle 
$\delta =8\pi Gm$ where $4Gm<1$~\cite{DJ}.
When $\varepsilon=-1$,
another coordinate transformation 
\begin{eqnarray}
r=\frac{c}{|\Lambda|^{1/2}\sin(2 c\ln{R})} 
~~{\rm and}~~\theta=\Theta 
~~(e^{k\pi/4c} < r <e^{(k+1)\pi/4c}~~{\rm and}~~
c^2=8GM,~k\in  {\rm Z})
\end{eqnarray}
results in the exterior region of the Schwarzschild 
type BTZ black hole~\cite{BTZ} with missing information of 
the point particle mass $m$ in Eqs.~(\ref{beq}) and (\ref{phieq}):
\begin{eqnarray}
ds^2=(|\Lambda|r^2-8GM)dt^2 -\frac{dr^2}{|\Lambda|r^2-8GM}-r^2d\theta^2.
\end{eqnarray}
As expected, the BTZ solution is one of general anti-de Sitter
solutions, of which physical meaning was not considered in Ref.~\cite{DJ}.
Note that the dimension of $m$ and $M$ has the 
square of mass in (3+1)D because it
represents the mass density per unit length along the string direction.

Here we want to solve Einstein equations with both a global string 
source and constant negative cosmological vacuum energy density.
We take a complex scalar field $\phi$ with Lagrange density
\begin{eqnarray}
{\cal L}= -\frac{1}{16\pi G}(R+2\Lambda)
+\frac{1}{2}g^{\mu\nu}\partial_\mu \overline \phi\partial_\nu \phi 
-\frac{\lambda}{4}(\overline\phi\phi -v^2)^2.
\end{eqnarray}
This model admits a string solution of the form 
\begin{eqnarray}
\phi=|\phi|(r)e^{in\theta}.
\end{eqnarray}
{}For the cylindrically symmetric configurations, the Euler-Lagrange
equations read under the metric in Eq.~(\ref{cyl}):
\begin{eqnarray}
&&\frac{1}{r}\frac{d N}{d r} =8\pi G \Big( \frac{d |\phi|}{dr}\Big)^2 \\
&&\frac{1}{r}\frac{dB}{dr}=2|\Lambda|
- 8\pi G\biggl\{ B\Bigl(\frac{d|\phi|}{dr}\Bigl)^{2}+\frac{n^2}{r^2}|\phi|^2
 +\frac{\lambda}{2}(|\phi|^2-v^2)^2\biggr\}\\
&&\frac{d^{2}|\phi|}{dr^{2}}+\Bigl(\frac{dN}{dr}+\frac{1}{B}\frac{dB}{dr} 
+\frac{1}{r}\Bigr)\frac{d|\phi|}{dr}=
\frac{1}{B}\Bigl(\frac{n^2|\phi|}{r^2}
+\lambda (|\phi|^2-v^2)|\phi|\Bigr).
\end{eqnarray}
Though we will consider the spacetime with horizons,
we concentrate only on the regular configurations connecting  
$|\phi|(0)=0$ and $|\phi|(\infty)=v$ smoothly.
By asking the reproduction of Minkowski spacetime in 
the limit of no matter ($T^{\mu}{}_\nu=0$),
and zero vacuum energy ($\Lambda\rightarrow 0$), we can choose an appropriate
set of boundary conditions, $B(0)=1$ and $N(\infty)=0$. 
The Einstein equation for the metric function $B(r)$ 
is then expressed in terms of scalar amplitude
\begin{eqnarray}
B(r)&=& 
\exp\Big[-8\pi G \int_r^\infty dr' r' \Big(\frac{d|\phi|}{dr'}\Big)^2\Big] 
\Bigg\{ 2|\Lambda|\int^{r}_{0}  dr' r' 
\exp\Big[8\pi G \int^\infty_{r'} dr'' r''\Big(\frac{d|\phi|}{dr''}\Big)^2\Big]
\\ \nonumber 
 &&\hspace{0mm} -8\pi G\int^r_0 dr'r'
\exp\Big[8\pi G \int^\infty_{r'} dr'' r''\Big(\frac{d|\phi|}{dr''}\Big)^2\Big]
\bigg(\frac{n^2}{r'^2}|\phi|^2
 +\frac{\lambda}{2}(|\phi|^2-v^2)^2\bigg)
+  e^{N(0)}\Bigg\}. 
\end{eqnarray}
Under the approximation that $|\phi|=0$ for $r<r_{c}$ and 
$|\phi|=v$ for $r\ge r_{c}$, we read the possible form of 
$B(r)$ with the aid of constant $N(r)$ which is a valid approximation for 
$r\ge r_{c}$:
\begin{eqnarray}
\label{bform}
B(r)\approx |\Lambda|r^2
-8\pi G v^2 n^2 \ln r/r_{c}
-4\pi G v^2 n^2 +1.
\end{eqnarray}
The core radius $r_{c}=\sqrt{2}n/\sqrt{\lambda} v$ is determined 
by minimizing the core mass.
Another scale, the minimum point of $B(r)$, is 
$r_m\approx\sqrt{4\pi G v^2/|\Lambda|}n$, 
and it may coincide with the horizon scale $r_H$ when 
$2\pi Gv^2 \gg |\Lambda|/\lambda v^2$.
Note that the formation of a black hole is favored as the magnitude
of the cosmological constant 
scaled by the square of Higgs mass, $|\Lambda|/\lambda v^2$, 
becomes small and the symmetry breaking 
scale $v$ is large. 
The latter condition is obvious and the former condition can be understood by
the balance between the negative energy due to the negative cosmological
constant and positive matter contribution.
This energy balance also explains the reason why we can have 
the regular global $U(1)$ vortex solution 
in singularity-free curved spacetime.
Let us recall no-go theorem that this global $U(1)$ scalar model 
can not support finite-energy static regular vortex configuration
in flat spacetime,
so the global $U(1)$ vortex contains a logarithmic divergence
in its expression of the energy per unit length. 
This symptom can not be remedied by inclusion of 
(2+1)D gravity, namely a higher spin (spin 2) field, since
(2+1) dimensional Einstein gravity does not have propagating degree. 
Therefore, the global $U(1)$ vortex coupled to Einstein gravity leads to an
unavoidable physical curvature singularity~\cite{Gre}.
In the model of our consideration, the negative vacuum energy achieves a 
kind of energy balance in anti-de Sitter spacetime, and both the scalar
field and the curvature are regular everywhere even though we have 
coordinate singularity at the horizons.

{}From now on we will solve the field equation and Einstein equations,
and will show that there exist regular
global $U(1)$ vortex solutions in anti-de Sitter space and the base manifolds
of these configurations constitute smooth hyperbola, extremal black hole and
charged black hole.

Near the origin, $|\phi|(r)\sim \phi_0 r^n$ and the leading term of the power
series solution of $B(r)$ is given by
\begin{eqnarray}
B(r)\sim 1+ \Big(
\frac{|\Lambda|}{\lambda v^2}-2\pi Gv^2 
-\frac{8\pi G \phi_0^2}{\lambda v^2}\delta_{n1}\Big) 
(\sqrt{\lambda}vr)^2.
\label{beq0}
\end{eqnarray}
When the cosmological constant rescaled by the symmetry breaking scale
is smaller (larger) than the Planck scale, 
$B(r)$ starts to decrease (increase) near the origin. Despite of the
difficulty in systematic series expansion at the asymptotic region, the
leading term provides for sufficiently large $r$:
\begin{eqnarray}
|\phi|(r)&\sim&v-\frac{\phi_{\infty}}{r^{2}}\\
B(r)&\sim&|\Lambda|r^{2}-8\pi Gv^{2}n^{2}\ln{r/r_{c}}- 8G{\cal M}+1
+{\cal O}(1/r^{2}).\label{binf}
\end{eqnarray}
Here ${\cal M}$ is the integration constant and will be identified as 
the core mass of the global vortex. Since the form of Eq.~(\ref{binf})
is the same as that in Eq.~(\ref{bform}) and we know that 
the magnitude of two terms proportional to $G$ can be larger than
the sum of other two positive terms for some appropriate large $r$,
the existence of
horizons for the small magnitude of the negative cosmological constant can
easily be confirmed.

To elicit the above discussion explicitly, we compute the solitons by use of
numerical analysis.
Two regular $n=1$ vortex solutions connecting  $|\phi|(0)=0$ and 
$|\phi|(\infty)=v$ are illustrated in Figure~1.
When $|\Lambda|/\lambda v^2$ is large enough, i.e., 
the second term in the right hand
side of Eq.~(\ref{beq0}) is positive for small $r$ 
($|\Lambda|/\lambda v^2=1.0$ and $8\pi Gv^2=1.15$), $B(r)$ is monotonically 
increasing (See $(i-a)$ in Fig.~1).
When $|\Lambda|/\lambda v^2$ is intermediate,  
$B(r)$ has a positive minimum (See $(i-b)$ in Fig.~1).
Suppose that there exists a horizon $r_H$ such that $B(r_H)=0$, another
boundary condition has to be satisfied, which is obtained from the equation
for the scalar field
\begin{eqnarray}
\frac{d|\phi|}{dr}\bigg|_{r_H}= 
\frac{|\phi|(r_H)\Big[\frac{n^2}{r_H^2} +\lambda(|\phi|^2(r_H)-v^2)\Big]}{
8\pi Gr_H\Big[\frac{|\Lambda|}{4\pi G}
-\Big( \frac{n^2}{r_H^2}|\phi|^2(r_H)
+\frac{\lambda}{2}(|\phi|^2(r_H)-v^2)^2\Big)\Big]} .
\end{eqnarray}
Therefore, we solve the equations in one region with two boundaries for
a regular solution, in two regions with three  boundaries for an 
extremal black hole and
in three regions with four boundaries for a charged black hole.
Since the value of $B(r)$ also vanishes at the horizon $r_H$ for 
the extremal black
hole, the position of the horizon and the value of scalar field 
are obtained in
explicit forms:
\begin{eqnarray}\label{ehori}
r_H =\frac{n}{\sqrt{\lambda}v\left(
1-\sqrt{1-\frac{|\Lambda|}{2\pi G\lambda v^4} 
}\right)^{1/2}} 
~~{\rm and}~~
|\phi|_{H}=v\bigg(1 -\frac{|\Lambda| }{2\pi G \lambda v^4}\bigg)^{1/4}.
\end{eqnarray}
When a black hole is formed, an intriguing property should be mentioned:
The ratio of horizon scale and the core length 
($\sim r_{c}\sim 1/\sqrt{\lambda}v$)
tells us that the horizon lies outside the string core.
We investigate the extremal black hole configuration for various
($|\Lambda|/\lambda v^2$, $8\pi Gv^2$) values and find one for 
$|\Lambda|/\lambda v^2=0.1$ and $8\pi Gv^2=1.338$
(See $(ii)$ in Fig.~1). A Reissner-Nordstr\"{o}m type charged black hole
with two horizons at $r_H^{in}$ and $r_H^{out}$ is also obtained for 
$|\Lambda|/\lambda v^2=0.1$ and $8\pi Gv^2=1.4$
(See $(iii)$ in Fig.~1). 
%%%%%%%%%%%%%%%%%%%%%%%%%%%%%%%%%%%%%%%%%%%%%%%%%%%%%%%%%%%%%%%%%%%%%%
\begin{figure}
% GNUPLOT: LaTeX picture with Postscript
\setlength{\unitlength}{0.1bp}
\begin{picture}(3600,2160)(0,0)
\put(0,1100){\makebox(0,0)[c]{\large$B(r)$}}
\put(1800,-100){\makebox(0,0)[l]{\large$\sqrt\lambda v r$}}
\put(1541,291){\makebox(0,0){$\bullet$}}
\put(1541,391){\makebox(0,0){\large$r_H$}}
\put(1331,291){\makebox(0,0){$\bullet$}}
\put(1201,241){\makebox(0,0){\large$r_H^{in}$}}
\put(1821,291){\makebox(0,0){$\bullet$}}
\put(1991,241){\makebox(0,0){\large$r_H^{out}$}}
\put(1291,1353){\makebox(0,0){$(i\!\rm{-}\!{\em a})$}}
\put(1891,1053){\makebox(0,0){$(i\!\rm{-}\!{\em b})$}}
\put(2391,853){\makebox(0,0){$(ii)$}}
\put(2691,703){\makebox(0,0){$(iii)$}}
\put(3201,50){\makebox(0,0){6}}
\put(2728,50){\makebox(0,0){5}}
\put(2255,50){\makebox(0,0){4}}
\put(1782,50){\makebox(0,0){3}}
\put(1309,50){\makebox(0,0){2}}
\put(836,50){\makebox(0,0){1}}
\put(363,50){\makebox(0,0){0}}
\put(313,2060){\makebox(0,0)[r]{2.5}}
\put(313,1706){\makebox(0,0)[r]{2}}
\put(313,1353){\makebox(0,0)[r]{1.5}}
\put(313,999){\makebox(0,0)[r]{1}}
\put(313,645){\makebox(0,0)[r]{0.5}}
\put(313,291){\makebox(0,0)[r]{0}}
\end{picture}

\vskip 1em
\caption{Various metric functions: (i) regular hyperbola (two dotted lines),
(ii) extremal black hole with a horizon $r_H$ (solid line),
and (iii) charged black hole with two horizons at $r_H^{in}$ and
$r_H^{out}$ (dashed line).}
\label{figblst}
\end{figure}
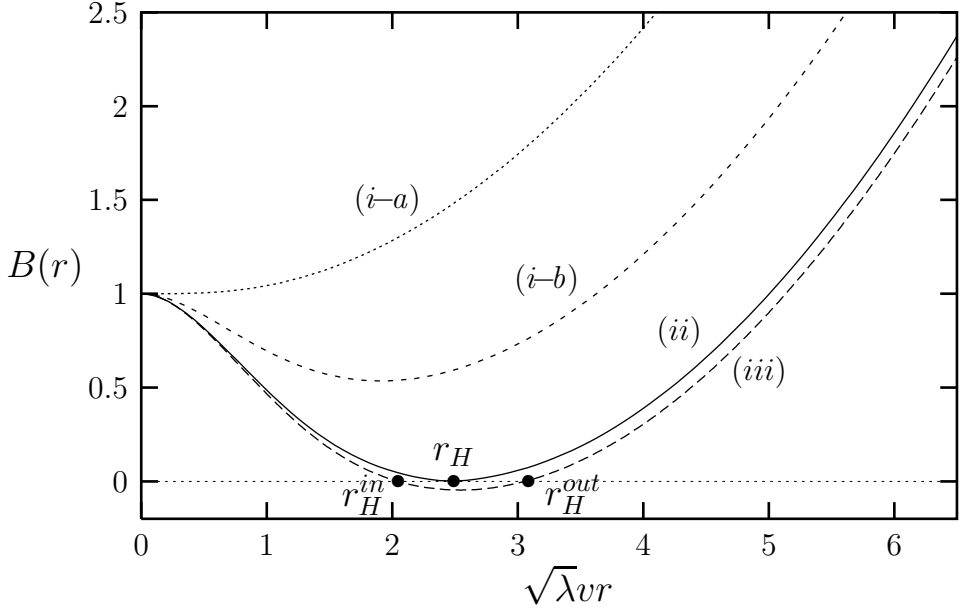
%%%%%%%%%%%%%%%%%%%%%%%%%%%%%%%%%%%%%%%%%%%%%%%%%%%%%%%%%%%%%%%%%%%%%%

Although we have a singularity at each horizon,
it is not physical singularity but coordinate artifact. 
This can be checked by reading the (2+1)D Kretschmann scalar:
\begin{eqnarray}
\lefteqn{R_{\mu\nu\rho\sigma}R^{\mu\nu\rho\sigma}
=4G_{\mu\nu}G^{\mu\nu}\label{curv}}&&\\
&=& %\hspace{-10mm}
4\mbox{Tr}
\bigg[\mbox{diag}\Big(-\frac{1}{2r}\frac{dB}{dr},-\frac{1}{2r}\frac{dB}{dr}
-\frac{B}{r}\frac{dN}{dr}, -\frac{1}{2}\frac{d^{2}B}{dr^{2}}-\frac{3}{2}
\frac{dB}{dr}\frac{dN}{dr}-B\frac{d^{2}N}{dr^{2}}-B\Big(\frac{dN}{dr}\Big)^{2}
\Big)^2\bigg].\nonumber 
\end{eqnarray}
Since both $N(r)$ and $B(r)$ are regular everywhere 
for all configurations of our consideration, no divergent curvature
appears at the horizon, and for the regular scalar configurations that
behave $|\phi|(r)\sim r^n$ for small $r$, the Kretschmann scalar
is also finite at the origin.
Therefore, unlike the global $U(1)$ strings which 
include unavoidable curvature singularity in the spacetime with 
zero cosmological constant~\cite{Gre}, 
the string configurations obtained in anti-de Sitter space
have no such curvature singularity.
{}Furthermore, since a possible divergent curvature at the core
of global  vortex is tamed by
the regular behavior of scalar amplitude $|\phi|(r)$,
no divergence of curvature appears, which is encountered in the charged black
hole formed due to infinite electric self energy by a point charge~\cite{BTZ}.

Now we look into the planar motions of massive and massless test particles
orthogonal to the string direction, which are described by the geodesic 
equations:
\begin{eqnarray}
\label{radial}
\frac{1}{2}\Big(\frac{dr}{ds}\Big)^2&=& -\frac{1}{2}\bigg(
B(r)\Big(m^2 +\frac{L^2}{r^2}\Big)
-\frac{\gamma^2}{e^{2N(r)}}\bigg)
\end{eqnarray}
with two constants of motion, $\gamma$ and $L$, associated with 
two Killing vectors $\partial/\partial t$ and 
$\partial/\partial\theta$ respectively.
{}From Eq.~(\ref{radial}), the elapsed coordinate time $t$ of a
test particle which moves from $r_0$ to $r$
is finite for regular solutions and becomes
infinite when it approaches to a horizon ($r\rightarrow r_H$). 
As we  expected, 
the spacetime with horizons depicts that of a black hole. 
The radial motion of a massless test particle ($m=0$ in Eq.~(\ref{radial}))
is unbounded for 
$\gamma\ne 0$. On the other hand, 
any massive particle ($m=1$ in Eq.~(\ref{radial})) can 
never escape the black hole irrespective of values of $\gamma$ and $L$, since
the asymptotic space is hyperbolic. 
The size of the boundary is determined as a function of 
$\gamma$ and $L$. 
Obviously, for the radial motion, the boundary $r_{\infty}$ of the massive
test particle becomes large as $\gamma$ increases, 
and then the ratio $r_{\infty}/r_H$ is much larger than one.
Details of all possible planar geodesic motions 
and related physical implication 
will be provided in Ref.~\cite{KKK}.

Under the metric in Eq.~(\ref{cyl}) 
the conserved quasilocal mass per unit length measured by the static observer
is~\cite{brown}:
\begin{eqnarray}
8GM_q=2e^{N(r)}\Big(\sqrt{(|\Lambda|r^2+1)B(r)}-B(r)\Big).
\end{eqnarray}
Though the obtained spacetime is not asymptotically flat and thereby $M_{q}$
is not identified by Arnowitt-Deser-Misner mass, we compute it for
sufficiently large $r$
\begin{eqnarray}\label{adm}
M_q\stackrel{r\rightarrow\infty}{\longrightarrow}\pi n^{2}v^{2}
\ln r/r_{c} + {\cal M}.
\end{eqnarray}
The first logarithmically divergent term comes from the topological sector of
the long range Goldstone degree, and the second finite one is the core mass of
the global $U(1)$ vortex, which coincide with those in flat spacetime.

If we compare the obtained spacetime of the global vortices with that of the
electric field of a point charge~\cite{BTZ}, 
we can easily find a similarity between them
except for the core region. The reason why we have this resemblance can be
explained by duality transformation \cite{KL}. The dual transformed theory
equivalent to (2+1)D (or (3+1)D) global vortex model of our consideration is
written in terms of a dual vector (or second rank antisymmetric tensor) field
$A_{\mu}$ (or $A_{\mu\nu}$):
\begin{eqnarray} \label{Max}
Z&=&\int [g^{\frac{3}{4}}dg_{\mu\nu}][|\phi|^{-2}d|\phi|][dA_{\mu}][d\Theta]
\exp\biggl\{i\int d^{3}x\sqrt{g}\Bigl[-\frac{1}{16\pi
G}(R+2\Lambda)\nonumber\\
&&\hspace{10mm}+\frac{1}{2}g^{\mu\nu}\partial_{\mu}|\phi|\partial_{\nu}|\phi|
-V(|\phi|)
-\frac{v^{2}}{4|\phi|^{2}}g^{\mu\nu}g^{\rho\sigma}F_{\mu\rho}F_{\nu\sigma}
+\frac{v\epsilon^{\mu\nu\rho}}{2\sqrt{g}}F_{\mu\nu}\partial_{\rho}\Omega\Bigr]
\biggr\},
\end{eqnarray} 
where $F_{\mu\nu}=\partial_{\mu}A_{\nu}-\partial_{\nu}A_{\mu}$, and $\Omega$
is the topological sector of the Goldstone degree. 
{}For the cylindrically symmetric strings with $\Theta=n\theta$,
the last term of the Lagrangian in Eq.~(\ref{Max}) describes the 
charged point source coupled minimally to a dual gauge field $A_\mu$.
The kinetic term of this $A_\mu$ is the Maxwell term for large $r$
($|\phi|\stackrel{r\rightarrow\infty}{\longrightarrow}v$),
but the nonpolynomial interaction ($\sim 1/|\phi|^2
\stackrel{r\rightarrow 0}{\sim}1/r^{2n}$) plays the role of ultraviolet cutoff
to remove the possible curvature divergence at the origin.
Through this reformulation the role of the
topological charge $n$ in the original formulation is transmuted to the
electric charge $n$ in the dual transformed theory.

Brief comments about physical implication of these black cosmic strings are in
order. Let us consider a universe with tiny negative cosmological constant, 
e.g., our present universe with $|\Lambda| \le 10^{-83}{\rm GeV}^2$. 
{}From Eq.~(\ref{ehori}), the 
characteristic scale to distinguish regular strings from black strings is 
about $v\sim 0.3{\rm eV}$. 
This means that favorite form of survived global strings in anti-de Sitter
space is black hole type where the magnitude 
of the cosmological constant is about
the lower bound of the present universe.
Estimation of the horizon size in Eq.~(\ref{ehori}) gives us 
$r_H\sim 10^{6}{\rm pc}$ for the grand unification scale 
($v\sim 10^{15}{\rm GeV}$), $r_H\sim 10^{-2}{\rm A.U.}$ for 
the electroweak scale ($v\sim 10^2{\rm GeV}$). 
Though it is estimated in a perfect presumed
toy environment, this property that the horizon of GUT scale 
black cosmic string is larger than the diameter of our galaxy
($\sim 5\times 10^4{\rm pc}$) may imply some incompatibility between the black 
cosmic string produced in such early universe and the extremely small magnitude
of negative cosmological constant. 
Again, let us emphasize that the scales obtained above are 
the outcome of three energy scales of big difference, which are 
the Planck scale (the largest energy scale), the bound of vacuum energy
(the smallest measured scale in cosmology) and an intermediate symmetry
breaking scale.
In this sense, the obtained black string configurations in Fig.~1
seem to be unphysical since $v\sim (10^{18}\sim10^{19}){\rm GeV}$ and
$\lambda\sim 10^{-122}$ for $|\Lambda|\sim 10^{-83}{\rm GeV}^2$
and $1/\sqrt{G}\sim 10^{19}{\rm GeV}$.
Obtaining black string under the physical situation
seems beyond our numerical precision.
Even if the global cosmic strings are produced, 
the lifetime of a typical string loop is very
short due to the radiation of gapless Goldstone boson,
which is the dominant mechanism for energy loss~\cite{Dav}. 
The space outside the horizon of black cosmic string 
is almost flat except for tiny attractive force due to negative cosmological 
constant as shown in Eq.~(\ref{radial}) and Eq.~(\ref{binf}), 
and then the massless Goldstone bosons
can be radiated outside the horizon. 
However, almost all the energy accumulated inside 
the horizon remains eternally. 
Let us remind you that the physical radius of black cosmic
string is decided by the horizon which is usually much larger than the
size of normal cosmic string (the core radius $ \sim 1/\sqrt{\lambda} v$),
but the energy per unit length of both objects is the same as given
in Eq.~(\ref{adm}).
Therefore, the very existence of this horizon is expected to 
change drastically the physics related to the evolution of global $U(1)$ 
strings, e.g., the intercommuting of two strings or 
the production of wakes by moving long strings \cite{VS}.

\vspace{5mm}

The authors would like to thank A. Hosoya, H. Ikemori, R. Jackiw, Chanju Kim,
Hyung Chan Kim and Kimyeong Lee for  helpful discussions. This work was 
supported by the KOSEF(95-0702-04-01-3 and through CTP, SNU).

%%%%%%%%%%%%%%%%%%%%%%%%%%%%%%%%%%%%%%%%%%%%%%%%%%%%%%%%%%%%%%%%%%%%%%%%%%%
%%Macro for the bibliography%%%%%%%%%%%%%%%%%%%%%%%%%%%%%%%%%%%%%%%%%%%%%%%
%%%%%%%%%%%%%%%%%%%%%%%%%%%%%%%%%%%%%%%%%%%%%%%%%%%%%%%%%%%%%%%%%%%%%%%%%%%
\def\hebibliography#1{\begin{center}\subsection*{References
}\end{center}\list
  {[\arabic{enumi}]}{\settowidth\labelwidth{[#1]}
\leftmargin\labelwidth	  \advance\leftmargin\labelsep
    \usecounter{enumi}}
    \def\newblock{\hskip .11em plus .33em minus .07em}
    \sloppy\clubpenalty4000\widowpenalty4000
    \sfcode`\.=1000\relax}

\let\endhebibliography=\endlist
\begin{hebibliography}{100}
%%%%%%%%%%%%%%%%%%%%%%%%%%%%%%%%%%%%%%%%%%%%%%%%%%%%%%%%%%%%%%%%%%%%%%%%%%%%
\bibitem{VS} For a review, see A. Vilenkin and E.P.S. Shellard, {\it Cosmic
Strings and Other Topological Defects}, (Cambridge, 1994);
M.B. Hindmarsh and T.W.B. Kibble, Rept. Prog. Phys. {\bf 58} (1995) 477.
%%%%%%%%%%%%%%%%%%%%%%%%%%%%%%%%%%%%%%%%%%%%%%%%%%%%%%%%%%%%%%%%%%%%%%%%%%%%
\bibitem{BTZ} M. Ba\~{n}ados, C. Teitelboim and J. Zanelli,  Phys. Rev. Lett.
{\bf 69} (1992) 1849; M. Ba\~{n}ados, M. Henneaux, C. Teitelboim and 
J. Zanelli, Phys. Rev. D {\bf 48}  (1993) 1506.
%%%%%%%%%%%%%%%%%%%%%%%%%%%%%%%%%%%%%%%%%%%%%%%%%%%%%%%%%%%%%%%%%%%%%%%%%%%%
\bibitem{DJ} S. Deser and R. Jackiw, Ann. Phys. {\bf 153} (1984) 405 .
%%%%%%%%%%%%%%%%%%%%%%%%%%%%%%%%%%%%%%%%%%%%%%%%%%%%%%%%%%%%%%%%%%%%%%%%%%%%
\bibitem{Cle} G. Clement, Phys. Rev. D {\bf 50} 7119 (1994);
                    ~Phys. Lett. B {\bf 367} (1996) 70;
K.C.K. Chan and R.B. Mann, Phys. Rev. D {\bf 50} (1994) 6385,
Erratum-{\it ibid}~D {\bf 52} 2600;
J.S.F. Chan, K.C.K. Chan and R.B.  Mann, Phys. Rev. D {\bf 54} (1996) 1535.
%%%%%%%%%%%%%%%%%%%%%%%%%%%%%%%%%%%%%%%%%%%%%%%%%%%%%%%%%%%%%%%%%%%%%%%%%%%%
\bibitem{HS} A. Vilenkin and A.E. Everett, 
       Phys. Rev. Lett. {\bf 48} (1982) 1867;
E.P.S. Shellard, Nucl. Phys. B {\bf 283} (1987) 624;
D. Harari and P. Sikivie, Phys. Rev. D {\bf 37} (1988) 3438.
%%%%%%%%%%%%%%%%%%%%%%%%%%%%%%%%%%%%%%%%%%%%%%%%%%%%%%%%%%%%%%%%%%%%%%%%%%%%
\bibitem{Gre} R. Gregory, Phys. Lett. B {\bf 215} (1988) 663;
A.G. Cohen and D.B. Kaplan, Phys. Lett. B {\bf 215} (1988) 67;
G.W. Gibbons, M.E. Ortiz and F. Ruiz Ruiz, Phys. Rev. D {\bf 39} (1989) 1546.
%%%%%%%%%%%%%%%%%%%%%%%%%%%%%%%%%%%%%%%%%%%%%%%%%%%%%%%%%%%%%%%%%%%%%%%%%%%%
\bibitem{KKK} N. Kim, Y. Kim and K. Kimm, in preparation.
%%%%%%%%%%%%%%%%%%%%%%%%%%%%%%%%%%%%%%%%%%%%%%%%%%%%%%%%%%%%%%%%%%%%%%%%%%%%
\bibitem{brown} J.D. Brown and J.W. York, Phys. Rev. D {\bf 47} (1993) 1407;
                J.D. Brown, J. Creighton, and R.B. Mann,
                Phys. Rev. D {\bf 50} (1994) 6394.
%%%%%%%%%%%%%%%%%%%%%%%%%%%%%%%%%%%%%%%%%%%%%%%%%%%%%%%%%%%%%%%%%%%%%%%%%%%%
\bibitem{KL} Y. Kim and K. Lee, Phys. Rev. D {\bf 49} (1994) 2041;
K. Lee, Phys. Rev. D {\bf 49} (1994) 4265; 
C. Kim and Y. Kim, Phys. Rev. D {\bf 50} (1994) 1040.
%%%%%%%%%%%%%%%%%%%%%%%%%%%%%%%%%%%%%%%%%%%%%%%%%%%%%%%%%%%%%%%%%%%%%%%%%%%%
\bibitem{Dav} R.L. Davis, Phys. Rev. D {\bf 32} (1985) 3172; 
A. Vilenkin and T. Vachaspati, Phys. Rev. D {\bf 35} (1987) 1138; 
{}For a review, see Ref.~\cite{VS} and the references therein.
%%%%%%%%%%%%%%%%%%%%%%%%%%%%%%%%%%%%%%%%%%%%%%%%%%%%%%%%%%%%%%%%%%%%%%%%%%%%
\end{hebibliography}
\end{document}